\documentclass{aa}  

\usepackage{gensymb}
\usepackage{float} 
\usepackage{subcaption}
\usepackage{graphicx}
\usepackage{txfonts}
\usepackage[breaklinks=true]{hyperref}
\hypersetup{
    colorlinks=true,
    citecolor=blue,
    linkcolor=blue,
    filecolor=blue,      
    urlcolor=blue,
    }

\newcommand{\Mstar}{$M_{\rm \ast}$}
\newcommand{\Msol}{$M_{\rm \odot}$}  
\begin{document}

   \title{Uncovering the MIR emission of quiescent galaxies with \textit{JWST}}
   % MIRI Analysis of Massive Galaxies (MIRIAM galaxies)

  \author{David Bl\'anquez-Ses\'e\inst{1,2}
          \and
          G. E. Magdis\inst{1,2,3}
          \and
          C. Gómez-Guijarro\inst{4}
          \and
          M. Shuntov\inst{3}
          \and
          V. Kokorev\inst{5}
          \and
          G. Brammer\inst{3}
          \and
          F. Valentino\inst{6,1}
          \and
          T. D\'iaz-Santos\inst{7,8}
          \and
          E.-D. Paspaliaris\inst{9,10}
          \and
          D. Rigopoulou\inst{11,8}
          \and
          J. Hjorth \inst{12}
          \and
          D. Langeroodi\inst{12}
          \and
          R. Gobat\inst{13}
          \and
          S. Jin\inst{1,2}
          \and
          N. B. Sillassen\inst{1,2}
          \and
          S. Gillman\inst{1,2}
          \and
          T. R. Greve\inst{1,2}
          \and
          M. Lee\inst{1,2}
          }

   \institute{ Cosmic Dawn Center (DAWN), Denmark
   \and
            DTU Space, Technical University of Denmark, Elektrovej 327, DK-2800 Kgs. Lyngby,Denmark
   \and
            Niels Bohr Institute, University of Copenhagen, Jagtvej 128, DK-2200 Copenhagen, Denmark
   \and
             Universit\'e Paris-Saclay, Universit\'e Paris Cit\'e, CEA, CNRS, AIM, 91191, Gif-sur-Yvette, France
   \and     
            Kapteyn Astronomical Institute, University of Groningen, P.O. Box 800,
            NL-9700AV Groningen, The Netherlands
    \and
            European Southern Observatory, Karl-Schwarzschild-Str. 2, 85748 Garching, Germany
    \and
            Institute of Astrophysics, Foundation for Research and Technology-Hellas (FORTH), Heraklion, 70013, Greece
    \and
            School of Sciences, European University Cyprus, Diogenes street, Engomi, 1516 Nicosia, Cyprus
    \and
            National Observatory of Athens, Institute for Astronomy, Astrophysics, Space Applications and Remote Sensing, Ioannou Metaxa and Vasileos Pavlou, 15236 Athens, Greece
    \and
            Department of Astrophysics, Astronomy \& Mechanics, School of Physics, Aristotle University of Thessaloniki, 54124 Thessaloniki, Greece
    \and
            Astrophysics, Department of Physics, University of Oxford, Keble Road, Oxford, OX1 3RH, UK
    \and
             DARK, Niels Bohr Institute, University of Copenhagen, Jagtvej 128, 2200 Copenhagen, Denmark
    \and
            Instituto de Física, Pontificia Universidad Católica de Valparaíso, Casilla 4059, Valparaíso, Chile
            }

  %\abstract {In this work we present one of the first studies of the MIR emission of QGs beyond the local universe. We use deep JWST imaging in the SMACS 0723 cluster field in order to characterise their dust emission and assess the existence of PAHs. We make a selection QGs via their $UVJ$ colours and split it into two samples: a low-$z$ selection of 47 galaxies located at the cluster redshift $<z>\approx0.3$, and an intermediate-$z$ selection of 3 galaxies at z = 0.65, 0.67, 0.75. We then analyse their properties using the SED fitting code Stardust as well as a hybrid SED model based on local ellipticals\thanks{Publicly available at http://www.georgiosmagdis.com/software/}. First we observe a clear dichotomy in the MIR emission of QGs compared to that of SFGs. We find that the presence of dust in QGs beyond the local universe is not uncommon and also detect signatures of PAH emission at various strengths, pointing towards a plethora of different dust component emissions. A further characterisation of the relative strengths of PAH features reveals a low 7.7/11.3µm $\sim$ 1 intensity ratio, consistent with previous studies of local ellipticals and characteristic of soft radiation fields and evolved stellar populations.}
  \abstract{We present a study of the mid-IR (MIR) emission of quiescent galaxies (QGs) beyond the local universe. Using deep $JWST$ imaging in the SMACS-0723 cluster field we identify a mass limited ($M_{\ast} >10^{9}$\,M$_{\odot}$) sample of intermediate redshift  QGs ($0.2<z<0.7$) and perform modeling of their rest-frame UV to MIR photometry. We find that QGs exhibit a range of MIR spectra that are composed of a stellar continuum and a dust component that is 1-2 orders of magnitude fainter to that of star-forming galaxies. The observed scatter in the MIR spectra, especially at $\lambda_{\rm rest} > 5\,\mu$m, can be attributed to different dust continuum levels and/or the presence of Polycyclic Aromatic Hydrocarbons (PAHs) features. The latter would indicate enhanced 11.3- and 12.7\,$\mu$m PAHs strengths with respect to those at 6.2- and 7.7$\,\mu$m, consistent with the observed spectra of local ellipticals and indicative of soft radiation fields. Finally, we augment the average UV-to-MIR spectrum of the population with cold dust and gas emission in the far-IR/mm and construct a panchromatic UV-to-radio SED\thanks{Publicly available at http://www.georgiosmagdis.com/software} that can serve as a template for the future exploration of the interstellar medium of $z>0$ QGs with ALMA and $JWST$.}


   \maketitle

%-------------------------------------------------------------------
%-------------------------------------------------------------------

\section{Introduction}

Over the past decades, a population of massive galaxies with suppressed star formation, i.e quiescent galaxies (QGs), has been robustly established both photometrically \citep{Daddi_2005,Toft_2005,Kriek_2006} and spectroscopically \citep{Toft_2012,Whitaker_2013,DEugenio_2020,Valentino_2020} up to $z \sim$ 4-5. The recent launch of the \textit{James Webb Space Telescope} (\textit{JWST}) has allowed to spectroscopically confirm QGs up to $z$ = 4.68 \citep{Carnall_2023} and provided the opportunity to search for possible $z\sim$ 5 candidates \citep{Valentino_2023}. QGs are characterized by low levels of star formation compared to their main sequence star-forming counterparts \citep{Daddi_2007,Schreiber_2015} and red colours, as a consequence of their old and evolved stellar populations. 

A large volume of studies characterizing QGs across cosmic time has focused on their stellar properties \citep[e.g. ][]{Williams_2009,Tomczak_2014}, using optical/near-infrared (NIR) data. In addition, their far-infrared (FIR) emission, tracing cold dust and gas, has been been explored in the local and, more recently, in the distant universe  \citep[e.g.][]{Young_2011,Magdis_2021,Blanquez_2023}, primarily with the Atacama Large Millimeter Array (ALMA). 

However, their mid-infrared (MIR) regime has only been examined in the local universe \citep[e.g.][]{Bregman_2006,Kaneda_2008,Rampazzo_2013}, since the "limited'' sensitivity of MIR observations carried out until recently with e.g. the \textit{Spitzer Space Telescope}, has predominantly restricted the study of MIR emission to star forming galaxies (SFGs) and Active Galactic Nuclei (AGNs). In fact, most studies focusing on $z>0$ QGs intentionally select against galaxies with a detection in the MIR  
(e.g. MIPS 24$\,\mu$m), as the latter is commonly associated with emission of warm dust heated by  ongoing star formation or AGN activity. 

The situation has been radically transformed with \textit{JWST} that can reach sensitivities to comfortably extend the study of the  MIR spectra of QGs to higher redshifts through the  detection of dust continuum  emission and, if present, of Polycyclic Aromatic Hydrocarbons (PAHs) that can be used as a powerful tracer of  star formation / AGN activity \citep[e.g.][]{Foster_sch_2004,Pope_2008,Kirkpatrick_2015,Xie_2019}, of the ISM conditions \citep[e.g.][]{Galliano_2008,Rigopoulou_2021} and even of the molecular gas reservoir \citep{Cortzen_2019} of a galaxy. 

In this work, we utilise deep 
imaging data obtained with the Near Infrared Camera \citep[NIRCam;][]{Rieke_2005} and the Mid-Infrared Instrument \citep[MIRI;][]{Rieke_2015,Bouchet_2015} as part of the \textit{JWST} Early Release
Observations \citep{Pontopiddan_2022} towards the 
SMACSJ0723.3-7327 (SMACS-0723) cluster field. Using the same set of observations, \citet{Langeroodi_23} presented a detailed study of the MIR colors of low- to intermediate-redshift galaxies and AGNs, and showed how PAH features enable distinguishing between SFGs, QGs, and AGNs in NIRCam and MIRI mid-infrared colour-colour diagrams. Here, we focus on a sample of intermediate redshift QGs to explore their MIR spectra - a feat that until recently was  unattainable beyond the local universe - and  highlight the feasibility of future observations towards this direction.

% When excited by radiation in star-forming (SF) regions, polycyclic aromatic hydrocarbons (PAHs) give rise to characteristic emission features in the mid-infrared range, notably at 3.3, 6.2, 7.7, 8.6, and 11.3 μm. This emission may account for as much as 10%–20% of the infrared luminosity in starburst galaxies, and the 7.7 μm feature alone may account for 50% of the total PAH luminosity (Draine & Li 2007; Smith et al. 2007; Wu et al. 2010; Shipley et al. 2013). As such, PAH luminosity is an excellent star formation rate (SFR) indicator for galaxies in the nearby universe (e.g., Förster Schreiber et al. 2004; Treyer et al. 2010; Shipley et al. 2016; Xie & Ho 2019).
%At low redshifts (z  0.3), PAH emission-based mid-infrared color selections of SF galaxies (SFGs) and active galactic nuclei (AGNs) have been carried out using photometry from the Infrared Space Observatory (ISO) or the Spitzer Space Telescope (Laurent et al. 2000; Veilleux et al. 2009). Due to the limited sensitivity of Spitzer, observations of PAH features at higher redshifts (z ∼ 0.3–2.8) have been limited to bright (ultra) luminous infrared galaxies (LIRGs) and AGNs (Pope et al. 2008; Kirkpatrick et al. 2015).
%The Mid-Infrared Instrument (MIRI) on board JWST offers an unprecedented opportunity to study the properties of much fainter galaxies at mid-infrared wavelengths.

The layout of the paper is as follows: In Section 2 we describe the utilized data and the methodology we followed for the  construction of a multi-band photometric catalogue in the JWST SMACS-0723 field. In Section 3 we present the sample selection of QGs and the derivation of their MIR properties through the modeling of their UV to MIR broadband photometry. In Section 4 we present the results, explore the possibility of the presence of PAHs in the spectra of intermediate redshift QGs, and construct a panchromatic UV-to-radio SED template for the population. Finally, in section 5 we provide a summary of our main findings. Throughout this work, we assume a standard $\Lambda$CDM cosmology with $\Omega_{M}$ = 0.3, $\Omega_{\Lambda}$ = 0.7 and $H_{0}$ = 70 km s$^{-1}$ Mpc$^{-1}$, adopt the Chabrier initial mass function (IMF) \citep{Chabrier_2003} and the AB magnitude system \citep{Oke_1974}.

%--------------------------------------------------------------------
%-------------------------------------------------------------------

\section{Data}
In this section we describe the data used for this work consisting of $HST$ and $JWST$ observations of the SMACS-0723 cluster field, centered around a massive lensing cluster located at $z_{\rm cluster} = 0.387$. All the data considered here are publicly available and were retrieved from the Mikulski Archive for Space Telescopes (MAST). 

\subsection{\textit{HST} archival data}

The $HST$ data were obtained from the treasury programme Reionization Lensing Cluster Survey \citep[RELICS;][]{Coe_2019}. Observations were carried out in the F435W, F606W, F814W filters from the Advanced Camera for Surveys (ACS) and F105W, F125W, F140W, F160W filters from the Wide Field Camera 3 (WFC3), covering a wavelength range from 0.435$\,\mu$m to 1.6$\,\mu$m.

\subsection{\textit{JWST} data}
SMACS-0723 is one of the first target fields observed by \textit{JWST} in the Early Release Observations (ERO) program \citep[ID 2736;][]{Pontopiddan_2022} after the telescope commissioning. The photometric observations consist of data from the Near Infrared Camera \citep[NIRCam;][]{Rieke_2005}, the Near Infrared Imager and Slitless Spectrograph \citep[NIRISS;][]{NIRISS} and the Mid-Infrared Instrument \citep[MIRI;][]{Rieke_2015,Bouchet_2015}. The observations covering the wavelength range from 0.9$-$18$\,\mu$m were carried out with the F090W, F150W, F200W, F277W, F356W, F444W NIRCam filters; F115WN, F200WN NIRISS filters, and F770W, F1000W, F1500W, F1800W MIRI filters, with a total observing time of 6.2\,hrs in the MIRI bands. The NIRCam pointings, consisting of two adjacent fields of view, cover an area of 2.2' $\times$ 2.2' each. Embedded within that area the MIRI coverage is located, with an area of 112.5" $\times$ 73.5". Since the MIRI photometric filters are the essential component of this work, we only consider the data within the MIRI coverage for the rest of the analysis.

\begin{figure}[H]
    \centering
    \includegraphics[width =\linewidth]{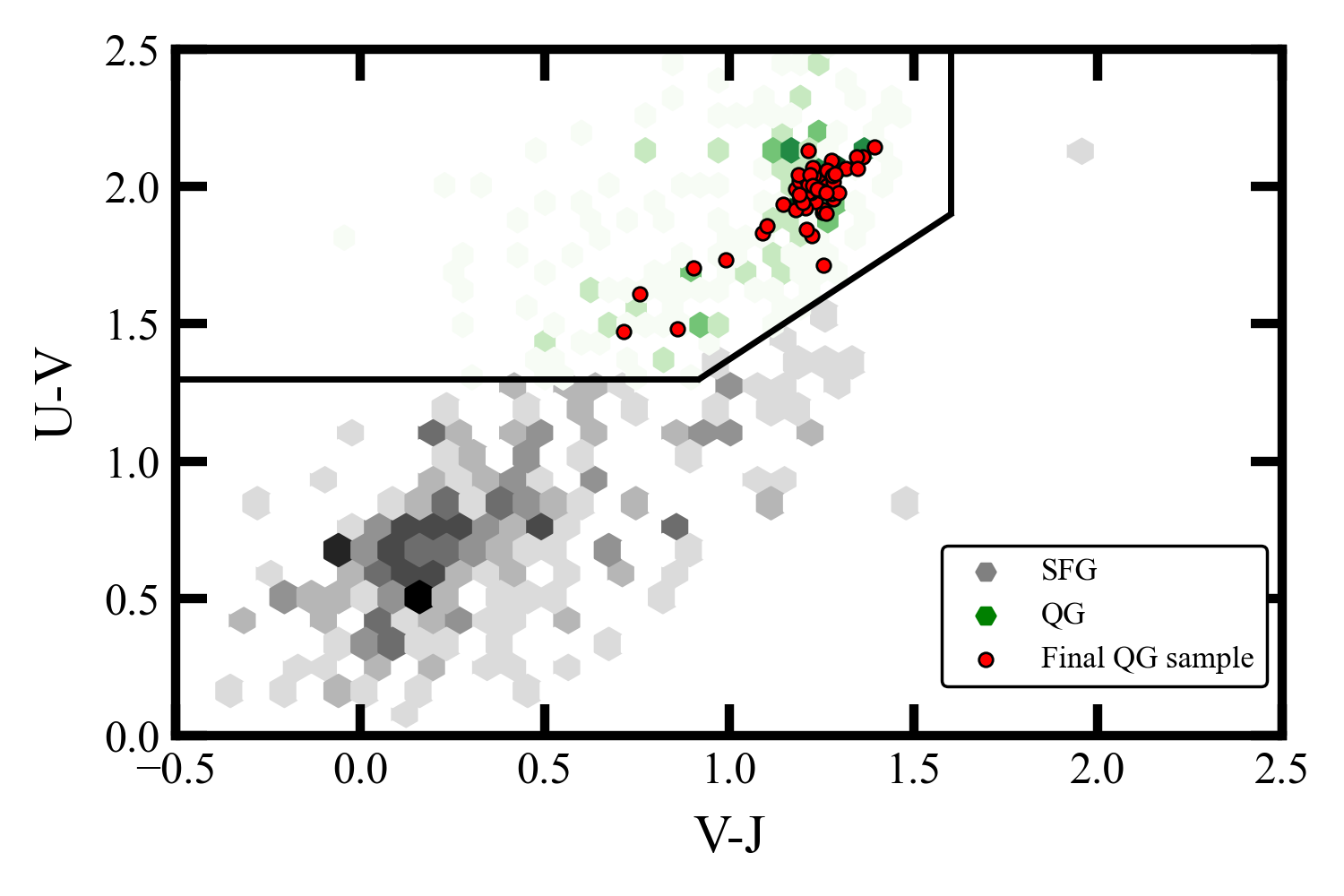}
    \caption{\textbf{UVJ diagram.} Rest-frame $U - V$, $V - J$ colour-colour diagram of the sources in the SMACS-0723 catalogue. The black solid lines represent the standard selection box defined in \citet{Schreiber_2015}. The green and grey hexagons linearly scaled from 0 to 7) show the density distribution of the full sample of QGs and SFGs respectively, while the red circles denote the position of final sample of QGs considered in this study.}
    \label{fig:UVJ diagram}
\end{figure}

\begin{figure}[H]
    \centering
    \includegraphics[width =\linewidth]{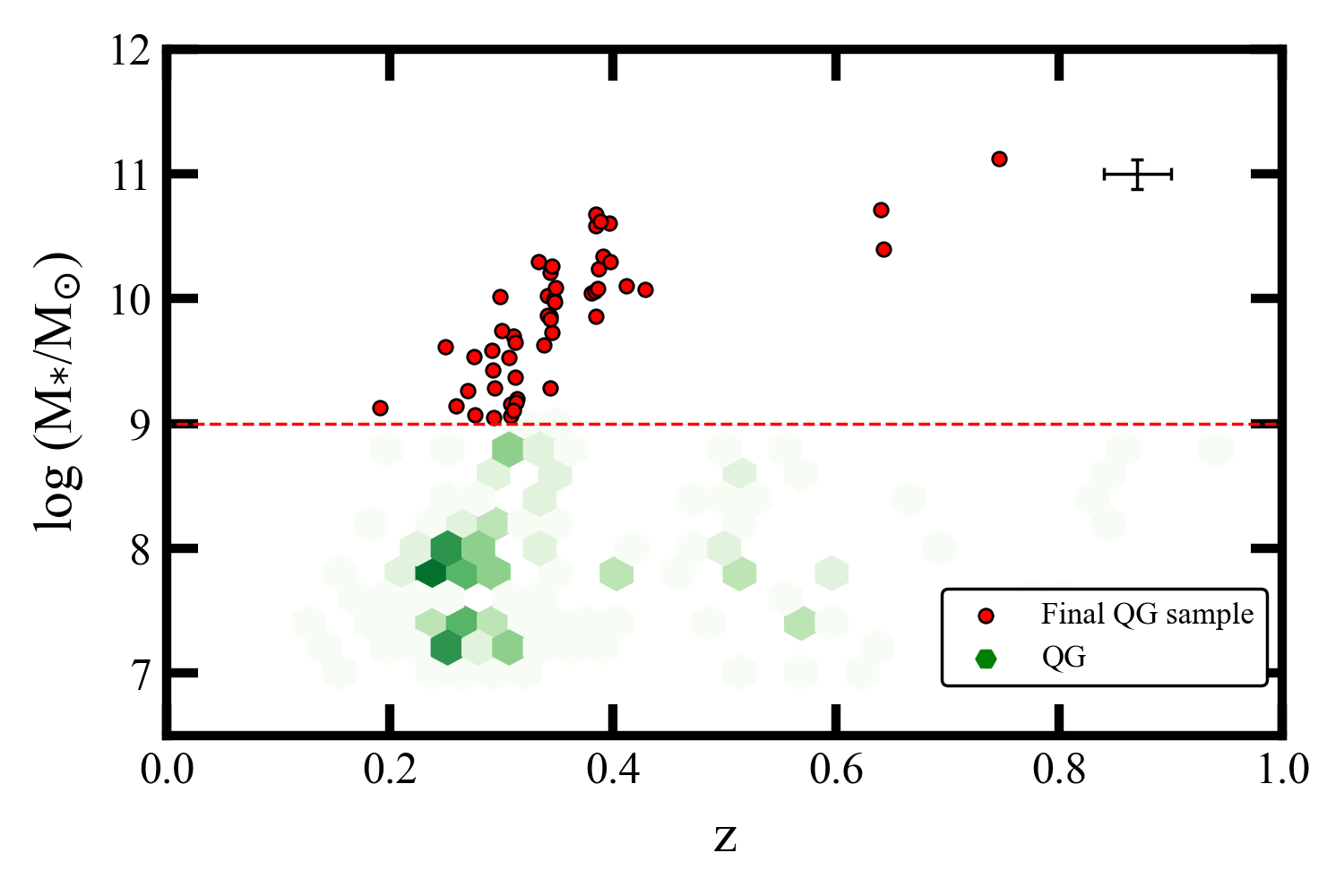}
    \caption{\textbf{Sample selection.} Distribution of sources in the redshift - stellar mass plane. The grey and green hexagons show the density distribution  of SFGs and QGs respectively. The horizontal line indicates the mass cut-off limit of log(M$_{*}$/M$_{\odot}) >$ 9 introduced in our selection. The red circles show the position of the final QGs sample, composed of 50 galaxies. On the top right we show representative errorbars for our catalogue. }
    \label{fig:selection}
\end{figure}
\subsection{Catalogue construction}
To reduce the NIRCam and MIRI data in the SMACS0723 field, we retrieved the level-2 products from MAST, and processed them with the \textit{grizli} pipeline \citep{grizli}. A similar data reduction methodology has been presented in e.g. \citet{Kokorev_2023,Valentino_2023} and will be fully described in Brammer et al. (in prep.). We ensure to give particular care to the photometric zero-points corrections relative to the \textit{jwst\_1041.pmap}. We additionally include corrections and masking to reduce the effects of stray light and cosmic rays. Our mosaics include the updated sky flats for all NIRCam filters. For MIRI data, we additionally construct our own flat fields, by utilising all available exposures in the field. Finally, we include all the available optical and near-infrared data in the Complete Hubble Archive for Galaxy Evolution (CHArGE; \citealt{Kokorev_2022}). We align the images to Gaia DR3 \citep{gaia-collaboration2021},  co-add, and finally drizzle them \citep{Fruchter_2002} to a 0.02'' pixel scale for the Short Wavelength (SW) NIRCam bands, and to 0.04'' for all the remaining \textit{JWST}/NIRCam, MIRI, and $HST$ filters.

 We then construct the photometric catalogue using {\fontfamily{lmtt}\selectfont SourceXtractor++} \citep{Sextractor_1,Sextractor_2}. {\fontfamily{lmtt}\selectfont SourceXtractor++} is a flexible model-fitting engine that does source detection and simultaneous model-fitting on a number of images of different photometric bands. We detect on a weighted mean of the long wavelength bands and fit S\'ersic models to all sources in all $HST$ and $JWST$ bands. The fitted models in each band are convolved with the corresponding PSF model. For the $JWST$ bands, PSF models are obtained with WebbPSF \citep{Perrin_2014}. This model-fitting approach offers the advantage of consistently measuring photometry in images of largely different resolutions, especially when combining ACS, NIRCam and MIRI, ranging from 0.4 $\mu$m to 18 $\mu$m.

To determine the redshifts and the rest frame colours of the detected sources we use {\fontfamily{lmtt}\selectfont EaZY-py} \citep{Brammer_2008}. We first search for spectroscopic reshifts ($z_{\rm spec}$) by crossmatching our galaxies with the MUSE spectroscopic catalogue of \citet{MUSE_zspec}, and assign $z_{\rm spec}$ to 31 out of 1725 galaxies. For the remaining sources we derive their photometric redshifts ($z_{\rm phot}$)  by running {\fontfamily{lmtt}\selectfont EaZY-py} with its default parameters, setting the redshift range to 0 < $z$ < 18 with a z$_{\rm step} = 0.01\times{(1+z)}$. We utilize the "$corr\_sfhz\_13$" subset of models within {\fontfamily{lmtt}\selectfont EaZY-py}, which make use of redshift dependent star formation histories and dust attenuation. To assess the quality of the derived photometric redshifts we repeat the {\fontfamily{lmtt}\selectfont EaZY-py} run for the 31 sources that have $z_{\rm spec}$, letting this time the redshift as a free parameter. We find an excellent agreement between the derived $z_{\rm phot}$ and the $z_{\rm spec}$, without a systematic offset and a scatter of $\sigma$=0.02.  

To derive the physical properties of the galaxies, and in particular their stellar mass ($M_{\rm *}$),  we use {\fontfamily{lmtt}\selectfont FAST} \citep{Kriek_2009}. We adopt the $z_{\rm phot}$ estimates from {\fontfamily{lmtt}\selectfont EaZY-py} and fit  the available photometry for each source up to 4.4\,$\mu$m with the stellar population models of \citet{bruzual_2003}, an exponentially declining star formation history, fixed solar metallicity and a \citet{Calzetti_2000} dust attenuation law. The derived stellar masses are correlated with  those obtained  by {\fontfamily{lmtt}\selectfont EaZY-py} with  a median offset of 0.25\,dex. 

\begin{table}[h]
    \centering
    
    \begin{tabular}{lccc}
    \hline
        $\lambda_{eff}$ [µm] & Filter & Depth$^a$\\ \hline
        7.52 & MIRI/F770W & 26.9 \\
        9.87 & MIRI/F1000W & 27.3 \\
        14.92 & MIRI/F1500W & 26.8 \\
        17.87 & MIRI/F1800W & 26.2 \\ \hline
    \end{tabular}
    \footnotesize{\\$^a$Computed from 2" circular apertures at 5$\sigma$.}
    \caption{List of magnitude depths in the MIR regime obtained by $JWST$ observations in the SMACS 0723 field.}
    \label{table:depth_comp}
\end{table}

\begin{figure*}[h!]
    \centering
    \includegraphics[width = 1\linewidth]{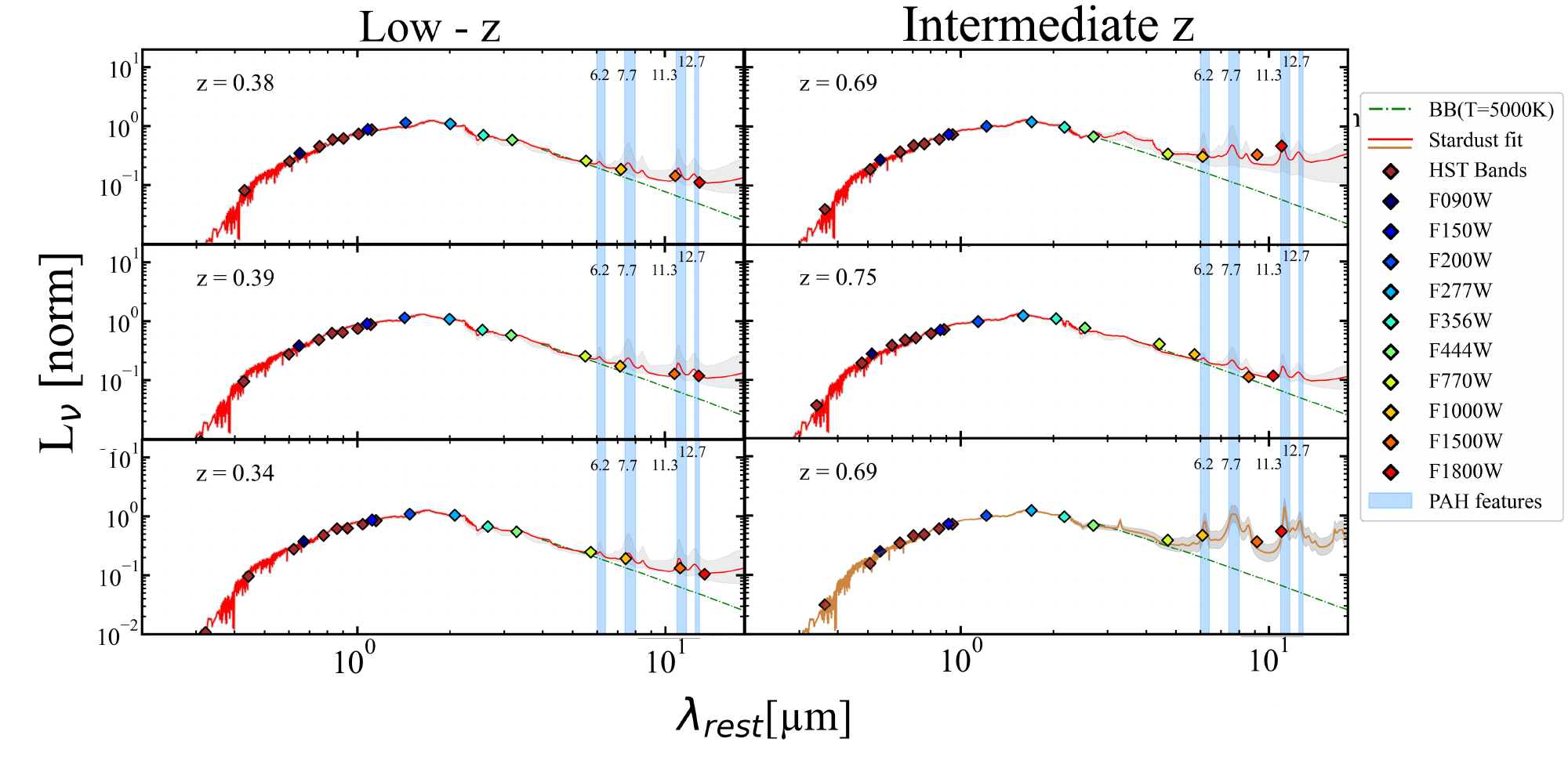}
    \caption{\textbf{SED fitting of QGs.} Rest frame SEDs of three low-$z$ (left panels) and the three intermediate-$z$ QGs (right panels) from our sample. The photometric data are color-coded according to the utilized broad-band filters while the solid lines (and the grey shaded regions) indicate the best fit models (and corresponding uncertainties). For galaxies where the MIR emission is best fit with the template of 
    \citet{Paspaliaris_2022} (P-QG) the best fit model is depicted with a red line. Instead the brown solid line indicates that the MIR emission of the galaxy is best fit with a DL07 model. The green dashed lines depict the  extrapolated stellar continuum emission modelled by a BB with $T$ = 5000\,K normalised at 3.5$\,\mu$m.. The light blue vertical shades indicate the positions and expected widths of the primary PAH features.}
    \label{fig:SED_highz}
\end{figure*}

\begin{figure*}[h!]
    \centering
    \includegraphics[width = 0.85\linewidth]{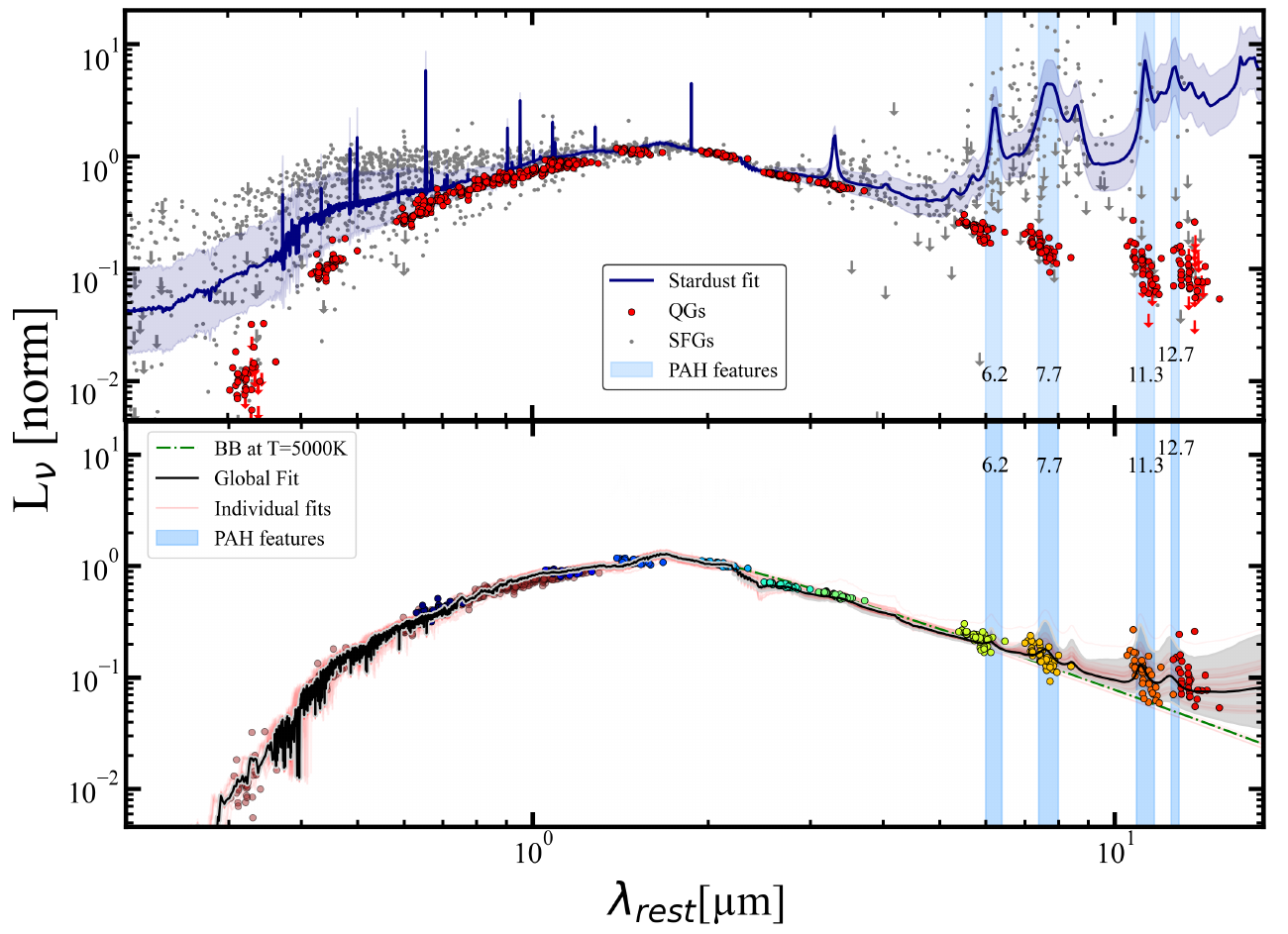}
    \caption{\textbf{Rest frame SEDs of SFGs and QGs in the SMACS catalogue.} \textit{Top:}  A compilation of the rest-frame SEDs of the SFGs in the SMACS catalogue (grey circles) and the QGs in our sample (red circles). All sources are normalised to their $K-$band luminosity.  The arrows depict the 3$\sigma$ upper limits. In dark blue we show the average {\fontfamily{lmtt}\selectfont Stardust} fit to the SFGs sample, with its associated uncertainty.  \textit{Bottom:} Rest frame SEDs of the low-$z$ subsample of QGs, normalised to their $K-$band luminosity. The photometric data are color-coded according to the utilized broad-band filters. The black solid line and the grey shaded area show the best fit model to the global data and its respective uncertainty, while the light red lines depict the fit to each individual galaxy in the sample. The green dashed line depicts the  extrapolated stellar continuum emission modelled by a BB with $T$ = 5000\,K normalised at 3.5$\,\mu$m. In both panels, the light blue vertical shades indicate the positions and expected widths of the primary PAH features.}
    \label{fig:SEDs}
\end{figure*}

In Table \ref{table:depth_comp}, we summarise the achieved 5$\sigma$ depths in the SMACS-0723 cluster field in the MIRI bands. In order to compute the mosaic sensitivities, we calculate the standard deviation of the measured flux densities in 3000 empty circular apertures that are randomly placed in the mosaic. Our analysis yields 5$\sigma$ magnitude depths of 26.9 and 26.2 at 7.7\,$\mu$m and 18\,$\mu$m respectively, 1-2 orders of magnitude deeper compared to some of the deepest  pre-$JWST$ MIR surveys (e.g. 23.1 and 19.6 at 8\,$\mu$m and 24\,$\mu$m with \textit{Spitzer} in the COSMOS field \citealt{COSMOS_2020,Jin_2018}). This along with the additional MIRI bands that sample the 8$-$24$\,\mu$m wavelength range provide a unique opportunity to explore the MIR emission of QGs beyond the local Universe.

%--------------------------------------------------------------------
%--------------------------------------------------------------------

\section{Analysis}
\label{sect:Analysis}
To select QGs from the parent sample we use the UVJ criterion \citep[e.g.][]{Williams_2009} after inferring the rest-frame colour from the best fit {\fontfamily{lmtt}\selectfont EaZY-py} SED models. For our purposes, we adopted the slightly modified
colour selection introduced by \citet{Schreiber_2015}:

\begin{equation}
\begin{cases}
U - V > 1.3,\\V-J < 1.6,\\U-V > 0.88 \times (V-J) + 0.49, 
\end{cases}
\end{equation}

This selection results in a sample of 240 QGs (see Fig. \ref{fig:UVJ diagram}) in the $0.12<z<2.35$ and $7.0<$ log$(M_{*}/M_{\odot})<11.4$ range (Figure \ref{fig:selection}). In an attempt to eliminate selection biases due to the different depths of available MIRI bands, we then introduce a stellar mass cut of \Mstar\ > 10$^{9}$\,\Msol, above which $\sim 95\%$ of the QGs are detected in both the F770W and F1000W bands at 5$\sigma$ significance (with the detection rate dropping to 75\% and 60\% in F1500W and F1800W respectively). The mass selection along with a visual inspection of the cutouts to discard artifacts and sources with poorly constrained photometry, yields a sample of 63 candidate QGs. Finally to identify possible dusty star forming galaxies (DSFGs) that are misclassified as QGs we fit the full available photometry of each source (0.43$-$ 18$\,\mu$m) with {\fontfamily{lmtt}\selectfont Stardust} \citep{Kokorev_2021}. This photometric fitting code can simultaneously and independently fit a stellar \citep{Brammer_2008}, AGN \citep{Mullaney_2011} and dust component \citep{Draine_2007,Draine_2014} to the data in order to reproduce the observed SED of a galaxy and obtain its fundamental optical and IR properties (e.g $M_{\rm *}$, $L_{\mathrm{IR}}$, $M_{\rm dust}$). We find that eight sources in our QGs sample have infrared luminosities ($L_{\rm IR}$) consistent with that of DSFGs ($L_{\rm IR}$ > 10$^{12}$\,$L_{\odot}$;\citet{Lonsdale_2006,Casey_2014}) and are subsequently removed from the sample. As shown in Figure \ref{fig:selection} our final sample consists of 50 QGs with log$(M_{*}/M_{\odot})> 9$ of which 47 are at $z\approx0.35$ ($0.2<z<0.4$; low-$z$ sample) and three are at $z = 0.65, 0.69, 0.75$ (intermediate-$z$ sample). We note that a significant fraction of QGs in our sample with $0.2<z_{\rm phot}<0.3$, i.e sources with $z_{\rm phot} < z_{\rm cluster}$,  are likely to be cluster galaxies for which {\fontfamily{lmtt}\selectfont EaZY-py} slightly underestimates their redshift. In fact, the majority of these sources have $z_{\rm phot}$  consistent to $z_{\rm cluster}$ within 1-2$\sigma$ of the $z_{\rm phot}$ uncertainty ($\langle\sigma\rangle \approx 0.06$). Placing all low$-z$ QGs at $z_{\rm cluster} = 0.387$ does not change the main results presented below.

 In order to model the MIR emission of the QGs in our final sample we re-run {\fontfamily{lmtt}\selectfont Stardust} with some modifications. First, in addition to the \cite{Draine_2007} dust models (DL07) utilized by {\fontfamily{lmtt}\selectfont Stardust}, we also include the dust emission (continuum + PAHs) of the empirical template (P-QG) presented in \citet{Paspaliaris_2022} that was constructed to reproduce the average IR emission of 229 local ellipticals. This step ensures that we will be able to capture any weak dust continuum and/or PAH emission that might be present in the spectrum of QGs and which cannot be recovered by the DL07 dust models. Finally, we approximate the emission of the stellar templates at $\lambda_{\rm rest}>2.2\,\mu$m as a blackbody (BB) with a temperature $T$= 5000 K. Examples of the derived best fit SEDs for low-$z$ and intermediate-$z$ QGs are presented in Figure \ref{fig:SED_highz}.

\section{Results \& Discussion}
The fitting methodology described above indicates that the rest-frame MIR emission of nine 
galaxies from our sample can be fully reproduced by stellar emission (BB, $T$ = 5000\,K), 
without any contribution from a dust component. However, for the vast majority of the QGs 
(41$/$50) a dust component on top of the stellar emission is necessary to reproduce the 
observed MIRI fluxes, revealing the presence of non negligible amounts of dust in the 
interstellar medium of low- and intermediate$-z$ QGs with $M_{*} > 10^{9}$\,M$_{
\odot}$. We note that varying the adopted BB temperature for the stellar continuum within the range of $T=$ 3000-7000\,K does not affect our results.    

With the SED models of the QGs at hand, we can also compile the average rest-frame UV to mid-IR spectrum of the population in order to get a more global and informative picture of their rest-frame MIR emission and how it compares to that of SFGs and local ellipticals.  In Figure \ref{fig:SEDs} (top) we show the rest-frame SEDs of the QGs normalised to their rest frame K-band luminosity, as derived from the best fit {\fontfamily{lmtt}\selectfont Stardust} stellar component. We also include the rest-frame SEDs of all SFGs in our catalogue along with the average best fit model derived by {\fontfamily{lmtt}\selectfont Stardust}. As expected while there is a clear distinction in their UV-optical colour, there is also a mixture of the two populations in the UV-NIR part of the spectrum due to red DSFGs that mimic the colours of QGs. The dichotomy between the two population becomes  more striking in the MIR ($\lambda_{\rm rest}>5$\,$\mu$m) with QGs being up to two orders of magnitude fainter than the SFGs, for fixed stellar mass (K-band luminosity). A notable feature in the photometry of each MIRI band is a trend of decreasing flux density with increasing rest-frame wavelength, or equally with decreasing redshift. To investigate the origin of this anti-correlation we mimic our observations by considering a grid of model SEDs within the redshift range of our sample. Each SED was then convolved with the corresponding MIRI transmission curves in order to measure the synthetic photometry in each band, at each redshift. Bringing the synthetic photometry to rest-frame and normalising at $K-$band (similar to the procedure applied to the real data), we recover a similar negative gradient in each MIRI band to that observed in the real data. We thus conclude, that the perceived trend is produced by a combination of the redshift range of the QGs in our sample along with the shape of the throughput of the MIRI filters.

Focusing on the QGs, in Figure \ref{fig:SEDs} (bottom) we show the compilation of rest-frame SEDs of 
the low-$z$ QGs along with their individual fits and the best fit model to the running median of the cumulative data. 
While, on average, we can reproduce the  MIR emission  by a superposition of a stellar 
component and the P-QG template, it is clear that the scatter in the rest-frame MIR  
increases at longer wavelengths that correspond to the declining part of the stellar continuum emission. This indicates not only that dust is present in the ISM of low$-z$ QGs but also that the dust 
emission can vary significantly between sources. The diversity of the MIR emission of QGs is 
also evident  in the SEDs of the three intermediate-$z$ QGs presented in Figure 
\ref{fig:SED_highz};  while the UV-NIR regime of their SED is nearly identical, their MIR 
emission exhibit a range of different dust continuum levels and spectral features. This also reflects the diversity found in the Spitzer/IRS spectra of the nuclei of local ellipticals, that span from comprising  prominent PAH features (especially at 11.3- and 12.7$\,\mu$m) to featureless continuum emission \citep{Panuzzo_2011,Rampazzo_2013}.

While the scatter in the MIR emission of low and intermediate$-z$ QGs is likely to originate from different dust continuum levels, the fact that the most prominent PAHs features fall in, or close to the MIRI bands studied here, brings forward an alternative, and rather intriguing possibility. Namely, the scatter in the MIR emission of the QGs studied here could be driven by the presence and the variation of the intensity of PAHs in the MIR  spectra of QGs. While PAHs in QGs beyond the local universe are not yet directly detected, here we might be getting a first glance. If indeed this is the case then our QGs seem to have stronger 11.3- and 12.7\,$\mu$m PAHs with respect to those at 6.2 and 7.7$\,\mu$m. Interestingly, this would be consistent with the MIR spectra of local elliptical \citep{Bregman_2006,Kaneda_2008,Rampazzo_2013}, and in direct contrast to that of SFGs which exhibit much more prominent 6.2- and 7.7$\,\mu$m features \citep[e.g.][]{Joblin_2000,Galliano_2008}. Moreover, a low 7.7/11.3 PAH ratio has been associated to a predominance of neutral to ionized PAH molecules in the ISM \citep{Bregman_2006,Tielens_2008}. The latter can also be enhanced by a soft radiation field which provides less energetic photons to excite the shorter wavelength features, as well as to diminish the UV field strength and increase the ratio of neutral to ionized PAHs \citep{Bregman_2006,Rigopoulou_2021,Draine_21}. To obtain a rough quantitative estimate of the 7.7/11.3 ratio for our QGs we divide the median flux of the data points located at the selected wavelengths, yielding a ratio of 7.7/11.3 = 0.6$-$1 (depending on the adopted underlying continuum level). This value corresponds to an ionization parameter of $G_{0}/n_{e}\times(T_{gas}/10^{3}K)^{1/2}$ = 950 $\pm$ 75 cm$^{3}$ \citep{Galliano_2008}, fully consistent with a very soft radiation field, characteristic of the evolved stellar populations found in QGs \citep{Renzini_1998,Daddi_2000}. It is also worth mentioning that other mid-IR lines could also contribute to the broadband MIRI photometry.  For example, a fraction of the F1500W flux density could be attributed to the [NeII] 12.8$\,\mu$m, which is largely blended with the 12.7$\,\mu$m PAH feature. However, based on high spectral resolution observations of local ellipticals, the intensity of the 12.7$\,\mu$m feature is $\geq$ $\times3$ larger relative to that of the [NeII] 12.8$\,\mu$m emission line \citep{Panuzzo_2011}. Nevertheless, we stress that the presence, and subsequently the strength of PAHs and of other atomic or molecular species can only be confirmed through follow-up MIR spectroscopy.

\begin{figure}[H]
    \centering
    \includegraphics[width = \linewidth]{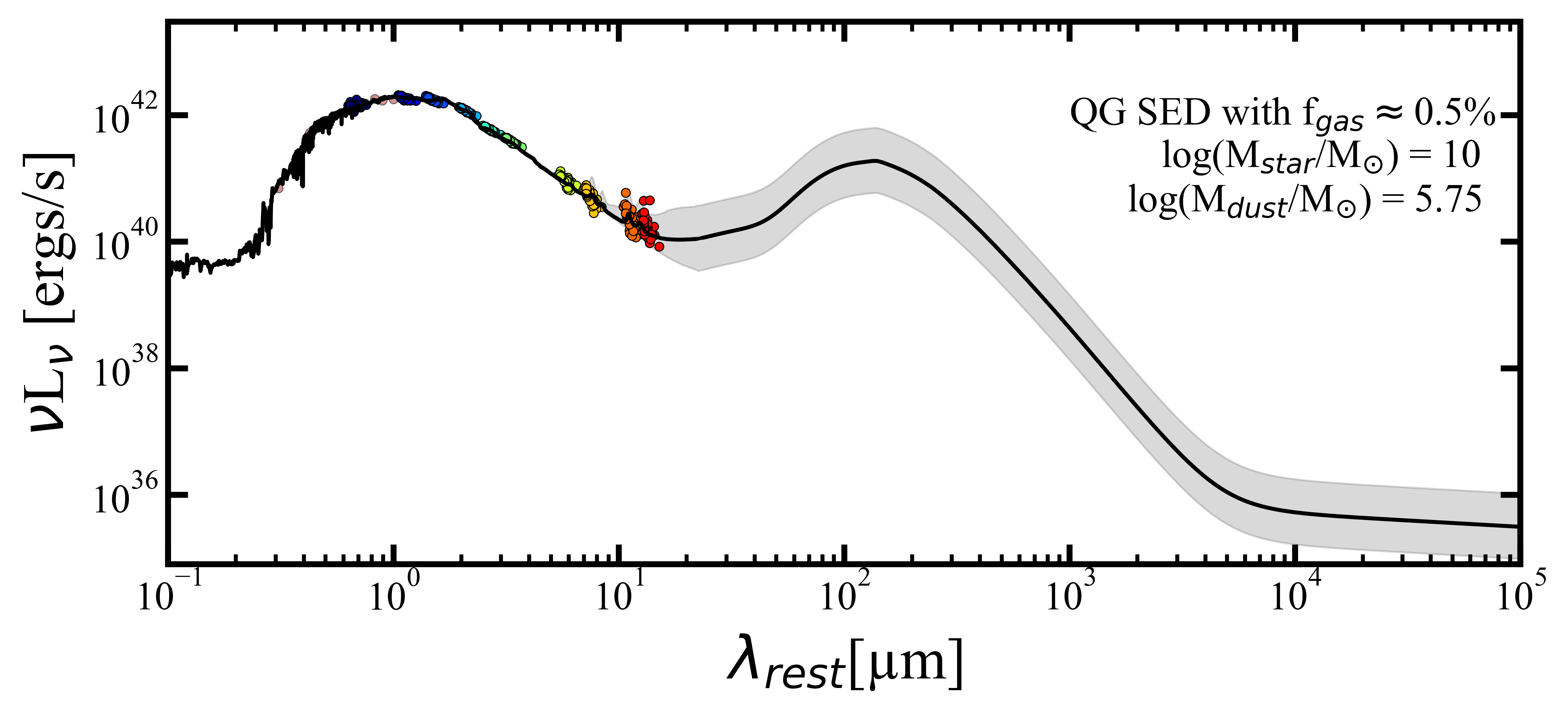}
    \caption{\textbf{Panchromatic template SED of  $z>0$ QGs.} Augmentation of the UV-to-MIR spectrum as obtained in this work with  the FIR SED model of QGs presented in  \citet{Magdis_2021}. The UV-to-MIR spectrum has been normalised to log(M$_{*}$/$M_{\odot}$) = 10 and the FIR model has been scaled to fit the MIR data resulting in a dust mass of log(M$_{\rm dust }$/$M_{\odot}$) = 5.75  and a gas fraction of $\approx$ 0.5\%. The full SED is shown as a solid black line and its associated uncertainty as a grey shaded region.}
    \label{fig:SED_panchromatic}
\end{figure}

Finally, we attempt to bridge our work with recent FIR studies that have charecterised the 
cold dust emission and the ISM mass budget of distant QGs. In particular we bring together 
the average UV-to-MIR model SED of the $z\approx 0.35$ QGs presented in Figure \ref{fig:SEDs}
(bottom), with the MIR-to-radio SED model introduced by \citet{Magdis_2021}, which is  
representative of the FIR emission of $z>0.3$ QGs. We first normalise the UV-to-MIR SED to 
log(M$_{*}$/$M_{\odot}$) = 10 and then conjoin the two models at 20$\,\mu$m by scaling the 
\citet{Magdis_2021} template. This scaling corresponds to a dust mass of 
log($M_{\rm dust}$/$M_{\odot}$) = 5.75 yielding a dust fraction of $f_{\rm dust} 
\approx 0.005\%$ and a gas fraction of $f_{\rm gas} \approx 0.5\%$, assuming a dust to gas 
mass ratio of 100. These values are consistent with the sharp decline of $f_{\rm gas}$ QGs 
between $z=2$ and the present day that are reported in the literature \citep[e.g.][]{Gobat_2018, Magdis_2021, Blanquez_2023}. Additionally, the QG model SED  can be used to derive an independent estimate of the ionization parameter. To this end we first scale the SED model to the mean stellar mass of the QG sample, i.e. log(M$_{*}$/$M_{\odot}$) = 10.5. We then employ the photodissociation region model described in \citet{Kaufman_1999,PDTtoolbox} and convert the total infrared luminosity of the model ($L_{\rm {IR}} = 1.4 \times 10^{8}\,L_{\odot}$), to a UV radiation field ($G_{0}$ = 12.5). Finally, assuming a range of gas temperatures ($T_{\mathrm{gas}} \approx 50-300$\,K) and electron densities ($n_{e} \approx 10^2 - 10^5$ cm$^{-3}$) appropriate for the $G_{0}$ value of the template, we infer an ionization parameter of $G_{0}/n_{e}\times(T_{gas}/10^{3}K)^{1/2}$ $\approx$ 1-1300 cm$^3$ that nicely brackets  our estimate based on 7.7/11.3 PAH interband ratio. The panchromatic, UV-to-radio, template SED is presented in 
Figure \ref{fig:SED_panchromatic} and is made publicly available\thanks{Publicly available at 
http://www.georgiosmagdis.com/software} to facilitate future studies of QGs beyond the local 
universe.

%--------------------------------------------------------------------
%--------------------------------------------------------------------

\section{Conclusions}
In this work we presented an initial  study of the MIR emission of  QGs beyond the local universe, taking advantage of deep NIRCam and MIRI $JWST$ observations of SMACS-0723 cluster field. We report the detection of dust emission in the MIR spectra ($\lambda_{\rm rest} = 4 - 14\,$µm) of $z=0.2-0.7$ QGs which is 1-2 order magnitudes fainter to that of SFGs. While for fixed stellar mass the rest-frame UV-to-NIR spectrum of the QGs appears to be rather homogeneous, the MIR emission of the population is characterised by a larger degree of diversification, especially at $\lambda_{\rm rest} > 5\,\mu$m. This scatter can be attributed to different dust continuum levels and/or the presence of PAHs in the spectra of QGs. The latter would indicate enhanced 11.3 and 12.7$\,\mu$m PAH features compared to those at 6.2 and 7.7\,$\mu$m, consistent with the observed MIR spectra of local ellipticals and the soft radiation fields that are expected in the ISM of passive galaxies. Finally, we construct and make publicly available a panchromatic (UV-to-radio) SED that could serve as a template for future studies of distant QGs. 

This study serves as a first step towards a better understanding of the MIR properties of QGs beyond the local universe and paves the way for follow-up spectroscopic observations that are necessary for the detailed characterisation of their ISM.

%--------------------------------------------------------------------
%--------------------------------------------------------------------

\begin{acknowledgements}
GEM, SG and DBS acknowledge financial support from the Villum Young Investigator grant 37440 and 13160 and the Cosmic Dawn Center (DAWN), funded by the Danish National Research Foundation under grant No. 140 PD. CGG acknowledges support from CNES.  DR acknowledges support from STFC through grant ST/W000903/1. JH and DL were supported by a VILLUM FONDEN Investigator grant (project number 16599). SJ is supported by the European Union’s Horizon Europe research
and innovation program under the Marie Skłodowska-Curie grant agreement No.
101060888. \end{acknowledgements}

\bibliographystyle{aa}
\bibliography{references.bib}

\end{document}